\documentclass{llncs}
\usepackage{pgfplotstable,filecontents}
\usepackage{tabularx}
\usepackage{amsmath}
\usepackage{amssymb}
\usepackage{wasysym}
\usepackage{booktabs}
\usepackage{makecell}
\usepackage{comment}
\pgfplotsset{compat=1.14}
\usepackage{graphicx}
\usepackage{authblk}
\usepackage[utf8]{inputenc}

\begin{document}

\title{Machine Learning for Large-Scale Quality Control of 3D Shape Models in Neuroimaging}

\author{Dmitry Petrov\thanks{these authors contributed equally}\inst{1, 2} \and Boris A. Gutman$^*$\inst{1} \and Shih-Hua (Julie) Yu\inst{1} \and Theo G.M. van Erp\inst{3} \and  Jessica A. Turner\inst{5} \and Lianne Schmaal\inst{23,24} \and Dick Veltman\inst{24} \and Lei Wang\inst{4} \and  Kathryn Alpert\inst{4} \and  Dmitry Isaev\inst{1} \and  Artemis Zavaliangos-Petropulu\inst{1} \and Christopher R.K. Ching\inst{1} \and  Vince Calhoun\inst{39} \and  David Glahn\inst{6} \and  Theodore D. Satterthwaite\inst{7} \and  Ole Andreas Andreasen\inst{8} \and  Stefan Borgwardt\inst{9} \and  Fleur Howells\inst{10} \and Nynke Groenewold\inst{10} \and  Aristotle Voineskos\inst{11} \and  Joaquim Radua\inst{12,33,34,35} \and  Steven G. Potkin\inst{3} \and  Benedicto Crespo-Facorro\inst{13,37} \and Diana Tordesillas-Gutiérrez\inst{13,37} \and  Li Shen\inst{14} \and  Irina Lebedeva\inst{15} \and  Gianfranco Spalletta\inst{16} \and  Gary Donohoe\inst{17} \and  Peter Kochunov\inst{18} \and Pedro G.P. Rosa\inst{19,32} \and Anthony James\inst{20} \and Udo Dannlowski\inst{25} \and Bernhard T. Baune\inst{30} \and André Aleman\inst{31} \and Ian H. Gotlib\inst{26}  \and Henrik Walter\inst{27} \and Martin Walter\inst{28,40,41} \and Jair C. Soares\inst{29} \and Stefan Ehrlich\inst{42} \and  Ruben C. Gur\inst{7} \and N. Trung Doan\inst{8} \and  Ingrid Agartz\inst{8} \and  Lars T. Westlye\inst{8,36} \and  Fabienne Harrisberger\inst{9} \and  Anita Riecher-Rössler\inst{9} \and Anne Uhlmann\inst{10} \and  Dan J. Stein\inst{10} \and  Erin W. Dickie\inst{11} \and  Edith Pomarol-Clotet\inst{12,33} \and Paola Fuentes-Claramonte\inst{12,33} \and Erick Jorge Canales-Rodríguez\inst{12,33,38} \and Raymond Salvador\inst{12,33} \and  Alexander J. Huang\inst{3} \and  Roberto Roiz-Santiañez\inst{13,37} \and  Shan Cong\inst{14} \and  Alexander Tomyshev\inst{15} \and  Fabrizio Piras\inst{16} \and Daniela Vecchio\inst{16} \and Nerisa Banaj\inst{16} \and Valentina Ciullo\inst{16}  \and Elliot Hong\inst{18} \and Geraldo Busatto\inst{19,32} \and Marcus V. Zanetti\inst{19,32} Mauricio H. Serpa\inst{19,32} \and Simon Cervenka\inst{21} \and Sinead Kelly\inst{22} \and Dominik Grotegerd\inst{25} \and Matthew D. Sacchet\inst{26} \and Ilya M. Veer\inst{27} \and Meng Li\inst{28} \and Mon-Ju Wu\inst{29} \and Benson Irungu\inst{29} \and Esther Walton\inst{42,43} \and Paul M. Thompson\inst{1}, for the ENIGMA consortium}

\institute{Imaging Genetics Center, Stevens Institute for Neuroimaging and Informatics, University of Southern California, Marina Del Rey, CA, USA
\and
The Institute for Information Transmission Problems, Moscow, Russia
\and
Department of Psychiatry and Human Behavior, University of California Irvine, Irvine, CA, USA
\and
Department of Psychiatry, Northwestern University, Chicago, IL, USA
\and
Psychology Department \& Neuroscience Institute, Georgia State University, Atlanta GA, USA
\and
Yale University School of Medicine, New Haven, CT, USA
\and
Department of Psychiatry, University of Pennsylvania School of Medicine, Philadelphia, PA, USA
\and
CoE NORMENT, KG Jebsen Centre for Psychosis Research, Division of Mental Health and Addiction, Oslo University Hospital \& Institute of Clinical Medicine, University of Oslo, Oslo, Norway
\and 
Department of Psychiatry, University of Basel, Basel, Switzerland
\and
MRC Unit on Risk \& Resilience to Mental Disorders, Department of Psychiatry and Mental Health, University of Cape Town, Cape Town, South Africa
\and
Centre for Addiction and Mental Health, Toronto, Canada
\and
FIDMAG Germanes Hospitalaries Research Foundation, Barcelona, Spain
\and 
University Hospital Marqués de Valdecilla, IDIVAL, Department of Psychiatry, School of Medicine, University of Cantabria, Santander, Spain
\and
Department of Radiology and Imaging Sciences, Indiana University School of Medicine, Indianapolis, IN, USA
\and
Mental Health Research Center, Moscow, Russia
\and
Laboratory of Neuropsychiatry, Santa Lucia Foundation IRCCS, Rome, Italy
\and 
School of Psychology, NUI Galway, Galway, Ireland
\and
Maryland Psychiatric Research Center,  University of Maryland School of Medicine, Baltimore
\and
Department of Psychiatry, Faculty of Medicine, University of São Paulo, São Paulo, Brazil
\and 
University of Oxford, Oxford, United Kingdom
\and
Centre for Psychiatry Research, Department of Clinical Neuroscience, Karolinska Institutet, Stockholm, Sweden 
\and 
Beth Israel Deaconess Medical Center, Harvard Medical School, Boston, MA, USA
\and
Orygen, The National Centre of Excellence in Youth Mental Health, Melbourne, Australia
\and 
Department of Psychiatry, VU University Medical Center, Amsterdam, The Netherlands
\and
Department of Psychiatry and Psychotherapy, University of M{\"u}nster, Germany
\and
Department of Psychology, Stanford University, Stanford, CA, USA
\and
Charité – Universitätsmedizin Berlin, corporate member of Freie Universität Berlin, Humboldt-Universität zu Berlin, and Berlin Institute of Health, Department of Psychiatry and Psychotherapy, CCM, Berlin, German
\and
Clinical Affective Neuroimaging Laboratory, Leibniz Institute for Neurobiology, Magdeburg, Germany
\and
University of Texas Health Science Center at Houston, Houston, TX, USA
\and 
Discipline of Psychiatry, Adelaide Medical School,
The University of Adelaide
\and
Interdisciplinary Center Psychopathology and Emotion regulation (ICPE), Neuroimaging Center (BCN-NIC), University Medical Center Groningen, University of Groningen, Groningen, The Netherlands
\and
Center for Interdisciplinary Research on Applied Neurosciences (NAPNA), University of São Paulo, São Paulo, Brazil
\and
CIBERSAM, Centro Investigación Biomédica en Red de Salud Mental, Barcelona, Spain
\and
Department of Clinical Neuroscience, Centre for Psychiatric Research, Karolinska Institutet, Stockholm, Sweden
\and
Department of Psychosis Studies, Institute of Psychiatry, Psychology and Neuroscience, King's College London, United Kingdom
\and
Department of Psychology, University of Oslo, Oslo, Norway
\and
CIBERSAM, Centro Investigación Biomédica en Red Salud Mental, Santander, Spain
\and
Radiology department, University Hospital Center (CHUV), Lausanne, Switzerland
\and
The Mind Research Network, Albuquerque, NM, USA
\and
Leibniz Institute for Neurobiology, Magdeburg, Germany
\and
Department of Psychiatry and Psychotherapy, University of Tübingen, Tübingen, Germany
\and
Division of Psychological and Social Medicine and Developmental Neurosciences, Faculty of Medicine, TU Dresden, Germany
\and
Psychology Department, Georgia State University, Atlanta, GA, USA
}

\maketitle              

\vspace{2mm}
\begin{abstract}
As very large studies of complex neuroimaging phenotypes become more common, human quality assessment of MRI-derived data remains one of the last major bottlenecks. Few attempts have so far been made to address this issue with machine learning. In this work, we optimize predictive models of quality for meshes representing deep brain structure shapes. We use standard vertex-wise and global shape features computed homologously across 19 cohorts and over 7500 human-rated subjects, training kernelized Support Vector Machine and Gradient Boosted Decision Trees classifiers to detect meshes of failing quality. Our models generalize across datasets and diseases, reducing human workload by 30-70\%, or equivalently hundreds of human rater hours for datasets of comparable size, with recall rates approaching inter-rater reliability.

\end{abstract}
\section{Introduction}
In recent years, large-scale neuroimaging studies numbering in the thousands and even 10’s of thousands of subjects have become a reality \cite{ENIGMAindiv}. Though automated MRI processing tools \cite{fischl2012freesurfer} have become sufficiently mature to handle large datasets, visual quality control (QC) is still required. For simple summary measures of brain MRI, QC may be a relatively quick process. For more complex measures, as in large studies of voxel- and vertex-wise features \cite{ENIGMAshape}, the QC process becomes more time-intensive for the human raters. Both  training of raters and conducting QC ratings, once trained, can take hours even for modest datasets. 
 
This issue is particularly relevant in the context of multi-site meta-analyses, exemplified by the ENIGMA consortium \cite{ENIGMAindiv}. Such studies, involving dozens of institutions, require multiple researchers to perform quality control on their cohorts, as individual data cannot always be shared. In addition, for meta-analysis studies performed after data collection, the QC protocols must be reliable in spite of differences in scanning parameters, post-processing, and demographics. In effect, QC has become one of the main practical bottlenecks in big-data neuroimaging. Reducing human rater time via predictive modeling and automated quality control is bound to play an increasingly important role in maintaining and hastening the pace of the scientific discovery cycle in this field. 

In this paper, we train several predictive models for deep brain structure shape model quality. Our data is comprised of the ENIGMA Schizophrenia and Major Depressive Disorder working groups participating in the ENIGMA-Shape project \cite{ENIGMAshape}. Using ENIGMA’s Shape protocol and rater-labeled shapes, we train a discriminative model to separate “FAIL”(F) and “PASS”(P) cases. For classification, we use a support vector classifier with a radial basis kernel (SVC) and Gradient Boosted Decision Trees (GBDT). Features are derived from the standard vertex-wise measures as well as global features. For six out of seven deep brain structures, we are able to reduce human rater time by 30 to 70 percent in out-of-sample validation, while maintaining FAIL recall rates similar to human inter-rater reliability. Our models generalize across datasets and disease samples.

\section{Methods}

Our goal in using machine learning for automated QC differs somewhat from most predictive modeling problems. Typical two-class discriminative solutions seek to balance misclassification rates of each class. In the case of QC, we focus primarily on correctly identifying FAIL cases, by far the smaller of the two classes \textbf{(Table \ref{table:data-overview})}. In this first effort to automate shape QC, we do not attempt to eliminate human involvement, but simply to reduce it by focusing human rater time on a smaller subsample of the data containing nearly all failing cases. Our quality measures, described below, reflect this nuance.

\subsection{MRI processing and shape features}

Our deep brain structure shape measures are computed using a previously described pipeline \cite{ShapeISBI,ShapeNature}, available via the ENIGMA Shape package. Briefly, structural MR images are parcellated into cortical and subcortical regions using FreeSurfer. Among the 19 cohorts participating in this study,  FreeSurfer versions 5.1 and 5.3 were used, depending on the institution. The binary region of interest (ROI) images are then surfaced with triangle meshes and parametrically (spherically) registered to a common region-specific surface template \cite{SphericalDemonsGutman}. This leads to a one-to-one surface correspondence across the dataset at roughly 2,500 vertices per ROI. Our ROIs include the left and right thalamus, caudate, putamen, pallidum, hippocampus, amygdala, and nucleus accumbens. Each vertex $p$ of mesh model $M$ is endowed with two shape descriptors:

	Medial Thickness, $D(p)=\Vert c_p - p\Vert$, where $c_p$ is the point on the medial curve $c$ closest to $p$. 

	$LogJac(p)$, Log of the Jacobian determinant $J$ arising from the template mapping,  $J:T_{\phi(p)}  M_t \to T_p M$.

Since the ENIGMA surface atlas is in symmetric correspondence, i.e. the left and right shapes are vertex-wise symmetrically registered, we can combine two hemispheres for each region for the purposes of predictive modeling. At the cost of assuming no hemispheric bias in QC failure, we effectively double our sample. 

The vertex-wise features above are augmented with their volume-normalized counterparts: $\{D,J\}_{normed}(p) = \frac{\{D,J\}(p)}{V^{\{\frac{1}{3},\frac{2}{3}\}}}$. Given discrete area elements of the template at vertex $p$, $A_t(p)$, we estimate volume as $V = \sum\limits_{p\in vrts(M)}^{}3A_t(p)J(p)D(p)$. We also use two global features: the shape-wide feature median, and the shape-wise 95th percentile feature threshold.

\subsection{Human quality rating}
\begin{figure} [ht]
   \begin{center}
   \begin{tabular}{c} 
   \includegraphics[height=4.5cm]{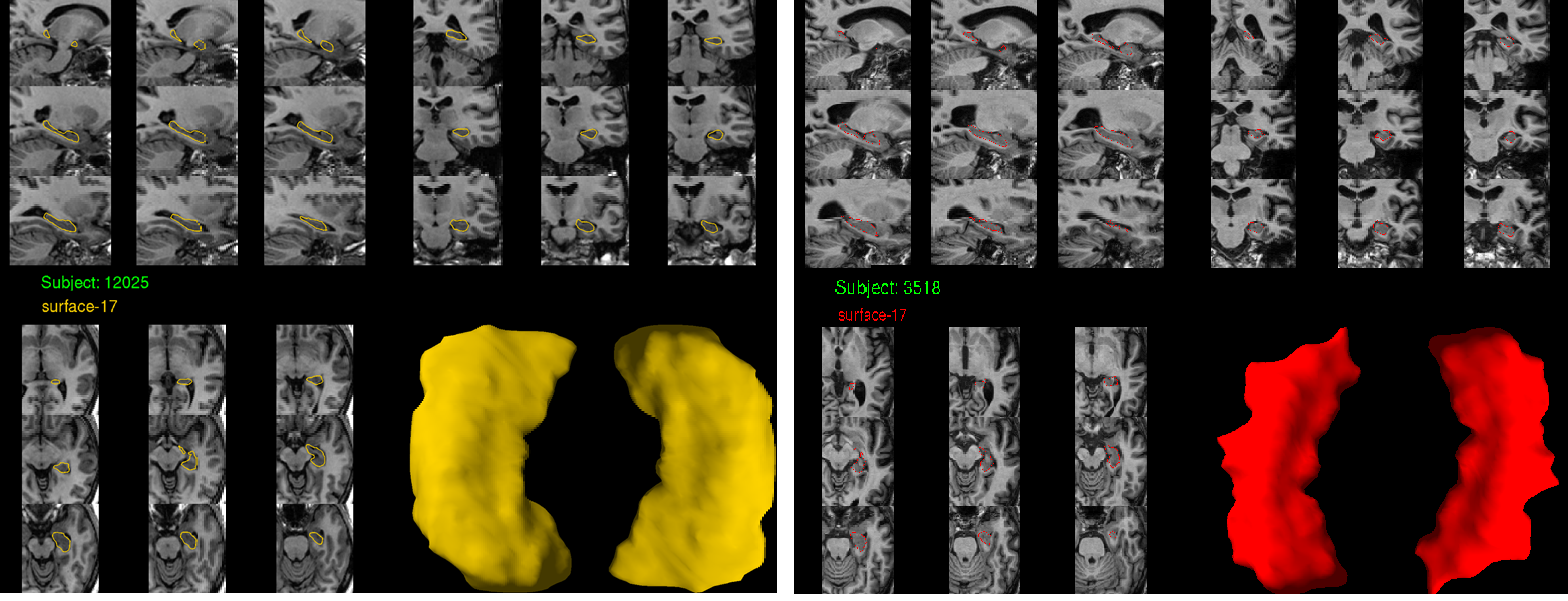}
   \end{tabular}
   \end{center}
   \caption 
   { \label{fig:QC_examples} 
\textbf{Example hippocampal shape snapshots used for human QC rating} 
\textbf{Left:} A mesh passing visual QC. \textbf{Right:} A mesh failing visual QC.}
   \end{figure}

Human-rated quality control of shape models is performed following the ENIGMA-Shape QC protocol\footnote{enigma.usc.edu/ongoing/enigma-shape-analysis}. Briefly, raters are provided with several snapshots of each region model as well as its placement in several anatomical MR slices (\textbf{Fig. \ref{fig:QC_examples}}). A guide with examples of FAIL (QC=1) and PASS (QC=3) cases is provided to raters, with an additional category of MODERATE PASS (QC=2) suggested for inexperienced raters. Cases from the last category are usually referred to more experienced raters for second opinions. Once a rater becomes sufficiently experienced, he or she typically switches to the binary FAIL/PASS rating. In this work, all remaining QC=2 cases are treated as PASS cases, consistent with ENIGMA shape studies. 

\subsection{Predictive models}

First, we used Gradient Boosted Decision Trees (GBDT). This is a powerful ensemble learning method introduced by Friedman \cite{gradboost} in which subsequent trees correct for the errors of the previous trees. In our experiments we used the Xgboost \cite{xgboost} implementation due to speed and regularization heuristics, with the logistic loss function. Second, we used Support Vector Classifier. Based on earlier experiments and the clustered nature of FAIL cases in our feature space, we used the radial basis  function (RBF) kernel in our SVC models. Indeed, in preliminary experiments RBF outperformed linear and polynomial kernels. We used scikit-learn's \cite{sklearn} implementation of SVC.

\subsection{Quality measures}

In describing our quality measures below, we use the following definitions. TF stands for TRUE FAIL, FF stands for FALSE FAIL, TP stands for TRUE PASS, and FP stands for FALSE PASS. Our first measure, $\textbf{F-recall} = \frac{TF}{TF + FP}$, shows the proportion of FAILS that are correctly labeled by the predictive model. The second measure, $\textbf{F-share} = \frac{TF + FF}{\text{Number of observations}}$, shows the proportion of the test sample labeled as FAIL by the model. 
Finally, we used a \textbf{modified F-score}, which allows us to compare models based on the specific requirements of our task, i.e. a very high F-recall and F-share substantially below 1, we use a variation on the standard F-score.
$$
\text{F-score\tiny{mod}}=2\times \frac{\text{F-recall}\times(1 - \text{F-share})}{\text{F-recall} + (1 - \text{F-share})}.
$$
Note that the modified F-score cannot equal 1, as in the standard case. An ideal prediction leads to \text{F-score\tiny{mod}} = 1 - \text{F-share}. The intuition behind our custom F-score is based on the highly imbalanced FAIL and PASS samples. A model that accurately labels all failed cases is only valuable if it substantially reduces the workload for human raters, a benefit reflected by \text{F-share}. 

\section{Experiments}

For each of the seven ROIs, we performed eight experiments defined by two predictive models (SVC and GBDT), two types of features (original and normed) and two cross-validation approaches. We tested "Leave-One-Site-Out" and 5-fold stratified cross-validation, as described below.  

\subsection{Datasets}

In our experiments, we used deep brain structure shape data from the ENIGMA Schizophrenia and Major Depressive Disorder working groups. 

\begin{table}
\resizebox{12.4cm}{!} {

\begin{tabular}{lllllllll}
\toprule
      &  FAIL \%            &    accumbens &      caudate &  hippocampus &     thalamus &      putamen &     pallidum &     amygdala \\
\midrule
Train & mean$\pm$std &  3.4$\pm$4.7 &  0.9$\pm$0.7 &  2.0$\pm$1.1 &  0.8$\pm$1.0 &  0.6$\pm$0.6 &  2.3$\pm$3.6 &  0.9$\pm$0.9 \\
      & max &         16.4 &          2.1 &          4.2 &          3.4 &          1.5 &         13.8 &          2.6 \\
      & min &          0.0 &          0.0 &          0.5 &          0.0 &          0.0 &          0.0 &          0.0 \\
      & size &        10431 &        10433 &        10436 &        10436 &        10436 &        10435 &        10436 \\
Test & mean$\pm$std &  4.7$\pm$4.5 &  1.4$\pm$1.5 &  4.9$\pm$4.8 &  1.4$\pm$1.5 &  0.4$\pm$0.8 &  1.9$\pm$2.0 &  0.8$\pm$0.9 \\
      & max &         10.5 &          3.5 &         11.4 &          3.5 &          1.6 &          3.8 &          2.1 \\
      & min &          0.0 &          0.0 &          0.0 &          0.0 &          0.0 &          0.0 &          0.0 \\
      & size &         3017 &         3018 &         3018 &         3018 &         3017 &         3018 &         3018 \\
\bottomrule
\end{tabular}
}
\\[10pt]
\caption{Overview of FAIL percentage mean, standard deviation, maximum and minimum for each site. Sample sizes for each ROI  vary slightly due to FreeSurfer segmentation failure.}
\label{table:data-overview}
\vspace{-2.5em}
\end{table}

	Our predictive models were trained using 15 cohorts totaling 5718 subjects' subcortical shape models from the ENIGMA-Schizophrenia working group. The ENIGMA-Schizophrenia (ENIGMA-SCZ) working group is comprised of over two dozen cohorts from around the world. The goal of the working group is to identify subtle effects of Schizophrenia and related clinical factors on brain imaging features. For a complete overview of ENIGMA-SCZ projects and cohort details, see \cite{ENIGMASCZ}. 

To test our final models, we used data from 4 cohorts in the Major Depressive disorder working group (ENIGMA-MDD), totaling 1509 subjects, for final out-of-fold testing. A detailed description of the ENIGMA-MDD sites and clinical questions can be found here \cite{ENIGMAMDD}.

\subsection{Model validation}

All experiments were performed separately for each ROI. The training dataset was split into two halves referred to as 'TRAIN GRID' and 'TRAIN EVAL.' The two halves contained data from each ENIGMA-SCZ cohort, stratified by the cohort-specific portion of FAIL cases. Model parameters were optimized using a grid search within 'TRAIN GRID', with either stratified 5-fold or Leave-One-Site-Out cross-validation. Parameters yielding the highest Area Under the ROC-curve were selected from among all cross-validation and feature types. 

Both SVC and GBDT produce probability estimates indicating the likelihood that the individual subject's ROI mesh is a FAIL case, $P_{FAIL}$. Exploiting this, we sought a probability threshold for each model selected during the grid search to optimize the modified F-score in the 'TRAIN EVAL' sample. This amounts to a small secondary grid search. To simplify traversing this parameter space, we instead sample \text{F-score\tiny{mod}} at regularly spaced values of $P_{FAIL}$, from 0.1 to 0.9 in 0.1 increments. This is equivalent to \text{F-share} in the 'TRAIN EVAL' sample (Eval \text{F-share},  \textbf{Table \ref{table:results-1}}).

Final thresholds (Thres in  \textbf{Table \ref{table:results-1}}) were selected based on the highest \text{F-score\tiny{mod}}, requiring that F-recall $\ge0.8$ - a minimal estimate of inter-rater reliability. It is important to stress that while we used sample distribution information in selecting a threshold, the final out-of-sample prediction is made on an individual basis for each mesh. 
\section{Results}
Trained models were deliberately set to use a loose threshold for FAIL detection, predicting 0.3-0.8 of observations as FAILs in the TRAIN GRID sample. These predicted FAIL observations contained 0.85-0.9 of all true FAILs, promising to reduce the human rater QC time by 20-70\%. These results largely generalized to the 'TRAIN EVAL' and test samples: \textbf{Table~\ref{table:results-1}} shows our final model and threshold performance for each ROI. 

With the exception of the thalamus, our final models' performance measures generalized to the test sample, in some cases having better sample F-recall and lower percentage of images still requiring human rating compared to the evaluation sample. A closer look suggests that variability in model predictions across sites generally follows human rater differences. \textbf{Table~\ref{table:results-2}} breaks down performance by test cohort. It is noteworthy that the largest cohort, M{\"u}nster (N = 1033 subjects, 2066 shape samples), has the best QC prediction performance. 
\begin{table}
\resizebox{12.4cm}{!} {
\begin{tabular}{llllrrrrrrr}
\toprule
        ROI & Model &  CV & Features & Thres &  \thead{Eval\\F-recall} &  \thead{Eval\\F-share} &  \thead{Eval\\F-score} &  \thead{Test\\F-recall} &  \thead{Test\\F-share} &  \thead{Test\\F-score} \\
\midrule
   Accumbens &  GBDT &  5-fold &    Normed &            0.014 &            0.83 &            0.2 &           0.81 &           0.75 &          0.27 &          0.74 \\
    Amygdala &  GBDT &  5-fold &    Normed &            0.007 &            0.89 &            0.4 &           0.72 &           0.76 &          0.40 &          0.67 \\
     Caudate &   SVC &    LOSO &  Original &            0.017 &            0.84 &            0.3 &           0.76 &           0.92 &          0.45 &          0.68 \\
 Hippocampus &  GBDT &  5-fold &    Normed &            0.010 &            0.85 &            0.2 &           0.83 &           0.91 &          0.29 &          0.80 \\
    Pallidum &  GBDT &  5-fold &    Normed &            0.009 &            0.86 &            0.2 &           0.83 &           0.86 &          0.30 &          0.77 \\
     Putamen &   SVC &    LOSO &    Normed &            0.007 &            0.84 &            0.4 &           0.70 &           0.65 &          0.44 &          0.60 \\
    Thalamus &   SVC &    LOSO &  Original &            0.007 &            0.84 &            0.4 &           0.70 &           1.00 &          1.00 &          0.00 \\
\bottomrule
\end{tabular}
}
\\[10pt]
\caption{Test performance of the models with the best \text{F-score\tiny{mod}} on evaluation (TRAIN EVAL). Excepting the thalamus, overall models' performance generalizes to out-of-sample test data}
\label{table:results-1}
\end{table}

At the same time, the "cleanest" dataset, Houston, with no human-detected quality failures, has the lowest \text{F-share}. In other words, Houston would require the least human rater time relative to its size, as would be hoped. 

\begin{figure}[!tbp]
  \centering
  \begin{minipage}[b]{0.49\textwidth}
    \includegraphics[width=\textwidth]{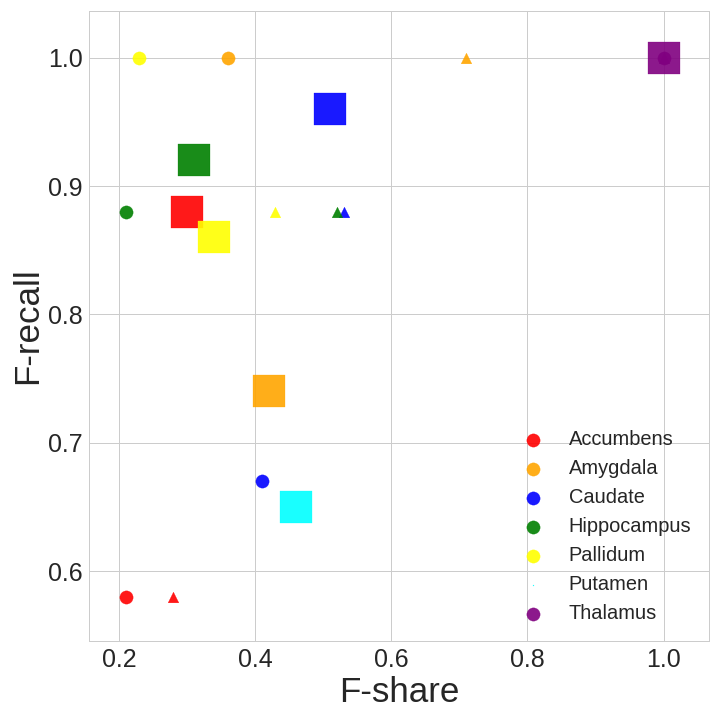}
    \end{minipage}
  \hfill
  \begin{minipage}[b]{0.49\textwidth}
    \includegraphics[width=\textwidth]{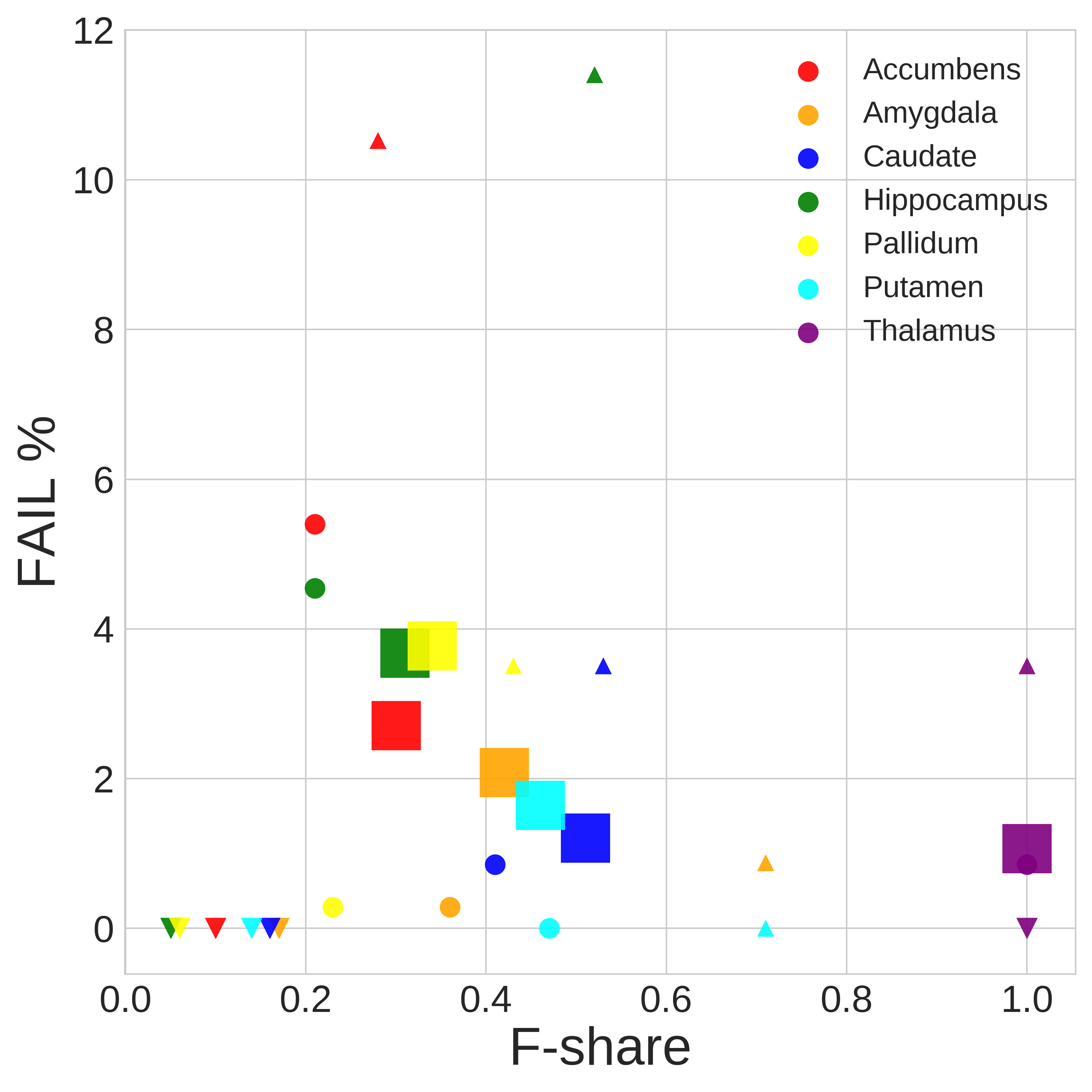}
    \end{minipage}
  \caption 
   { \label{fig:scatters} 
\textbf{Scatter plots of F-recall and actual FAIL case percentage vs. proportion of predicted FAIL cases on test datasets.} 
\textbf{Left:} F-share vs F-recall. \textbf{Right:} Fail F-share vs FAIL percentage. Mark size shows the dataset size. Mark shape represents dataset (site): \textbf{$\bigcirc$ - CODE-Berlin (N=176); $\Box$ - M{\"u}nster (N=1033);  $\bigtriangleup$ - Stanford (N=105); $\bigtriangledown$ - Houston (N=195)}}. 
  
\end{figure}

Visualizing the test results in \textbf{Figure \ref{fig:scatters}}, we see the trend for lower \text{F-share} with  higher overall dataset quality maintained by the smaller cohorts, but reversed by M{\"u}nster. This could be a reflection of our current models' bias toward accuracy in lower-quality data due to greater numbers of FAIL examples (i.e., FAILs in high and low quality datasets may be qualitatively different). At the same time, \text{F-recall} appears to be independent of QC workload reduction due to ML, with most rates above the $0.8$ mark.

\begin{table}
\begin{tabular}{lrrrrrrrr}
\toprule
         ROI & \thead{Berlin\\F-recall} &  \thead{Berlin\\F-share} & \thead{Stanford\\F-recall} &  \thead{Stanford\\F-share} & \thead{Munster\\F-recall} &  \thead{Munster\\F-share} & \thead{Houston\\F-recall} &  \thead{Houston\\F-share} \\
\midrule
   Accumbens &            0.58 &            0.21 &              0.58 &              0.28 &              0.88 &             0.30 &                - &             0.10 \\
    Amygdala &               1.00 &            0.36 &                 1.00 &              0.71 &              0.74 &             0.42 &                - &             0.17 \\
     Caudate &            0.67 &            0.41 &              0.88 &              0.53 &              0.96 &             0.51 &                - &             0.16 \\
 Hippocampus &            0.88 &            0.21 &              0.88 &              0.52 &              0.92 &             0.31 &                - &             0.05 \\
    Pallidum &               1.00 &            0.23 &              0.88 &              0.43 &              0.86 &             0.34 &                - &             0.06 \\
     Putamen &               - &            0.47 &                 - &              0.71 &              0.65 &             0.46 &                - &             0.14 \\
    Thalamus &               1.00 &            1.00 &                 1.00 &              1.00 &              1.00 &             1.00 &                - &             1.00 \\
\bottomrule
\end{tabular}
\\[10pt]
\caption{Performance of best models for each test site. Models are the same as in Table \ref{table:results-1}. Symbol '-' indicates that there were no FAILs for particular ROI and test site.}
\label{table:results-2}
\end{table}



\section{Conclusion}

We have presented a preliminary study of potential machine learning solutions for semi-automated quality control of deep brain structure shape data. Though some work on automated MRI QC exists \cite{EstebanMRIQC}, we believe this is the first ML approach in detecting end-of-the-pipeline feature failure in deep brain structure geometry. We showed that machine learning can robustly reduce human visual QC time for large-scale analyses for six out of the seven regions in question, across diverse MRI datasets and populations. Failure of the thalamus ML QC ratings to generalize out-of-sample may be explained by the region's specific features. Though we have only used geometry information in model training, MRI intensity, available to human raters for all ROI's, plays a particularly important role in thalamus ratings. The most common thalamus segmentation failure is the inclusion of lateral ventricle by FreeSurfer. Geometry is generally altered undetectably in such cases.  

Beyond adding intensity-based features, possible areas of future improvement include combining ML algorithms, exploiting parametric mesh deep learning, employing geometric data augmentation, and refining the performance measures. Specifically, mesh-based convolutional neural nets can help visualize problem areas, which can be helpful for raters. 

Very large-scale studies, such as the UK Biobank, ENIGMA, and others, are becoming more common. To make full use of these datasets, it is imperative to maximally automate the quality control process that has so far been almost entirely manual in neuroimaging. Our work here is a step in this direction. 



\section{Acknowledgements}
This work was funded in part by NIH BD2K grant U54 EB020403, Russian Science Foundation grant 17-11-01390 and other agencies worldwide.

%
%
\bibliography{Boris_refs2}
\bibliographystyle{splncs}

\end{document}